\documentclass[sigconf]{acmart}
\AtBeginDocument{%
  }

\usepackage{graphicx}
\usepackage{amsfonts}
\usepackage{amsmath}
\usepackage{multirow}
\usepackage{booktabs}
\usepackage{soul}
\usepackage{pifont}

\usepackage[normalem]{ulem}
\usepackage{bbding}
\usepackage{pifont}
\usepackage{tabularray}
\usepackage{enumitem}

\setcopyright{acmlicensed}
\copyrightyear{xxxx}
\acmYear{xxxx}
\acmDOI{XXXXXXX.XXXXXXX}
\acmConference[Conference acronym 'XX]{Make sure to enter the correct
  conference title from your rights confirmation email}{June 03--05,
  xxxx}{Woodstock, NY}
\acmISBN{978-1-4503-XXXX-X/2018/06}

\begin{document}

\title{Towards Blind Bitstream-corrupted Video Recovery: \\ A Visual Foundation Model-driven Framework}

\author{Tianyi Liu}
\affiliation{%
  \institution{School of EEE, Nanyang Technological University}
  \country{Singapore}}
\email{liut0038@e.ntu.edu.sg}

\author{Kejun Wu}
\affiliation{%
  \institution{School of EIC, Huazhong University of Science and Technology}
  \city{Wuhan}
  \country{China}}
\email{kjwu@hust.edu.cn}

\author{Chen Cai}
\affiliation{%
  \institution{School of EEE, Nanyang Technological University}
  \country{Singapore}}
\email{e190210@e.ntu.edu.sg}

\author{Yi Wang}
\affiliation{%
  \institution{Dept. of EEE, The Hong Kong Polytechnic University}
  \city{Hong Kong SAR}
  \country{China}}
\email{yi-eie.wang@polyu.edu.hk}

\author{Kim-Hui Yap}
\authornote{Corresponding Author}
\affiliation{%
  \institution{School of EEE, Nanyang Technological University}
  \country{Singapore}}
\email{ekhyap@ntu.edu.sg}

\author{Lap-Pui Chau}
\affiliation{%
  \institution{Dept. of EEE, The Hong Kong Polytechnic University}
  \city{Hong Kong SAR}
  \country{China}}
\email{lap-pui.chau@polyu.edu.hk}

\renewcommand{\shortauthors}{Liu et al.}

\begin{abstract}
Video signals are vulnerable in multimedia communication and storage systems, as even slight bitstream-domain corruption can lead to significant pixel-domain degradation. 
To recover faithful spatio-temporal content from corrupted inputs, bitstream-corrupted video recovery has recently emerged as a challenging and understudied task.
However, existing methods require time-consuming and labor-intensive annotation of corrupted regions for each corrupted video frame, resulting in a large workload in practice. 
In addition, high-quality recovery remains difficult as part of the local residual information in corrupted frames may mislead feature completion and successive content recovery.
In this paper, we propose the first blind bitstream-corrupted video recovery framework that integrates visual foundation models with recovery model, which is adapted to different types of corruption and bitstream-level prompts. 
Within the framework, the proposed Detect Any Corruption (DAC) model leverages the rich priors of the visual foundation model while incorporating bitstream and corruption knowledge to enhance corruption localization and blind recovery. 
Additionally, we introduce a novel Corruption-aware Feature Completion (CFC) module, which adaptively processes residual contributions based on high-level corruption understanding. 
With VFM-guided hierarchical feature augmentation and high-level coordination in a mixture-of-residual-experts (MoRE) structure, our method suppresses artifacts and enhances informative residuals.
Comprehensive evaluations show that the proposed method achieves outstanding performance in bitstream-corrupted video recovery without requiring a manually labeled mask sequence. 
The demonstrated effectiveness will help to realize improved user experience, wider application scenarios, and more reliable multimedia communication and storage systems.
\end{abstract}




\maketitle

\section{Introduction}
\label{sec:intro}

\begin{figure*}
  \centering
  \includegraphics[width=1\linewidth]{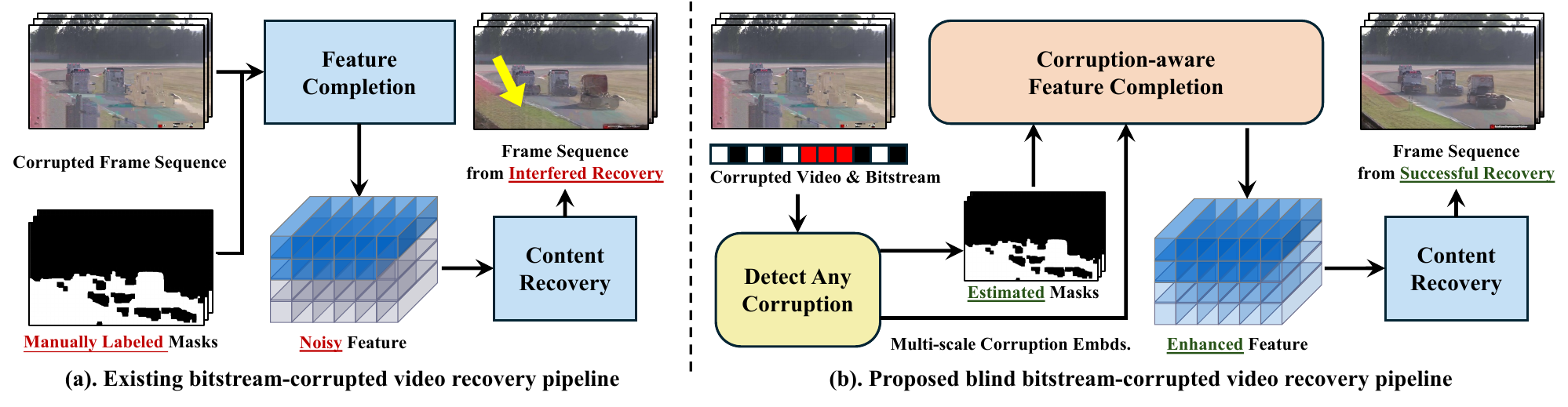}
  \vspace{-18pt}
  \caption{(a). Bitstream-corrupted video recovery framework requires manually labelled masks to identify the corrupted region, which incurs significant costs. (b). Blind bitstream-corrupted video recovery focuses on eliminating the need for mask annotation by effectively localizing the corrupted region and performing enhanced recovery.}
  \vspace{-1em}
  \label{fig:teaser}
\end{figure*}

With video becoming the dominant form of content in communication, entertainment and surveillance, high-quality playback is critical for modern multimedia systems~\cite{dobrian2011understanding, ciscoreport}.
Particularly, ensuring a seamless audiovisual experience is important for live-streaming and video-on-demand services, as well as for supporting vision-based multimedia systems such as monitoring~\cite{khanam2024yolov11, lu2025structureawaremotionadaptiveframework3d}, recognition~\cite{liu2022interactive}, and scene understanding~\cite{cai2025semantics, su2025annexe}.
However, it remains challenging, as bitstreams are highly vulnerable to corruption during both transmission and storage due to inter-frame correlation and heavy compression~\cite{stockhammer2003h, bross2011h265, qin2024byte,liu2023bitstream}.
Even minor packet loss or bit errors at the bitstream level can cause significant pixel-level degradation, often resulting in unpredictable and mixed artifacts~\cite{liu2023bitstream, willy2023most}.

To address this issue, traditional video restoration methods are not suitable because they are designed for specific global degradations~\cite{wang2019edvr, liang2024vrt}, such as compression artifacts~\cite{dong2015compression, lin2024toward}, noise~\cite{maggioni2021efficient, yue2020supervised}, or blur~\cite{chan2022generalization, sun2021deep}, which rely on task-specific priors and assume that the underlying content remains largely intact but globally distorted. 
This makes the methods ineffective in handling video corruption caused by bitstream errors, which often lack fixed priors and exhibit unpredictable distribution~\cite{liu2023bitstream}.
In recent years, video inpainting and error concealment methods have enabled partial general recovery by completely masking missing or damaged regions. 
However, their performance significantly degrades when dealing with large corrupted areas, making them unreliable for bitstream-corrupted video recovery.
To better address this issue, BSCVR~\cite{liu2023bitstream} is proposed as the baseline and state-of-the-art (SOTA) method that first takes advantage of residual textures, colors, and structures from corrupted regions to complete features in the embedding space before performing content recovery.
However, as shown in Fig.\ref{fig:teaser}(a), BSCVR and inpainting baselines typically rely on mask sequences that explicitly label corrupted regions in each frame, a requirement that is impractical for real-world scenarios due to the labor-intensive nature of frame-by-frame annotation. 
Despite this, the automatic detection and localization of video corruption remain unexplored. 
In addition, the diversity and complexity of corruption patterns pose significant challenges to self-attention-based feature completion~\cite{liu2023bitstream}, often resulting in suboptimal recovery quality. 
Indiscriminate use of residual information can further degrade performance; for instance, as shown in Fig.\ref{fig:teaser}, color-stripe artifacts may propagate during feature completion. 
Therefore, it is crucial to suppress noise and enhance informative residual features to approximate clean representations and improve recovery.

An important reason behind the aforementioned issues is that recovery models fail to establish a high-level understanding of the video corruptions encountered, thus lacking the prior knowledge needed to guide effective corruption detection and recovery.
Fortunately, the recent emergence of vision foundation models (VFMs) equipped with rich open-world knowledge offers a potential solution, as they demonstrate strong generalization capabilities across a wide range of visual tasks~\cite{sam, ravi2024sam2, caron2021dino, oquab2023dinov2, clip}.
To this end, we explore how to leverage the rich prior knowledge in VFM for more capable corruption detection and localization, with the aim of helping existing recovery models achieve blind bitstream-corrupted video recovery.
Specifically, we propose the Detect Any Corruption (DAC) module, which innovatively extracts video encoding knowledge embeddings and incorporates learnable auxiliary embeddings to perform cross-domain knowledge prompting to the image embeddings extracted from the foundational visual encoder. 
The improved perception from enhanced image embeddings helps decode more accurate masks, which is crucial for detecting and localizing video corruption effectively.
Furthermore, we design a novel Corruption-aware Feature Completion (CFC) module that can be simply embedded into existing recovery frameworks. 
In CFC, the video corruption feature is first enhanced at a macro level by incorporating multi-scale foundational corruption embeddings from VFM. 
Building on this, at an intermediate level, several residual experts are designed to dynamically collaborate with each other, leveraging the open world knowledge of VFM to achieve understanding of high-level corruption patterns and coordination of experts. 
Finally, at a micro level, each expert performs self-prompting to adaptively select the relevant and informative embeddings to refine and complete the overall features in detail.

Our main contributions can be summarized as follows.
\begin{itemize}[leftmargin=*, topsep=0pt]
\item We explore a novel research problem: \textbf{blind bitstream-corrupted video recovery} and present a VFM-driven blind video recovery framework. This task emphasizes realistic corruption modeling and overcomes the reliance on explicit corruption annotations simultaneously, making video recovery more authentic and deployable in real-world multimedia systems.
\item We propose Detect Any Corruption (DAC) to fill the gap in understanding, detecting, and localizing video corruption, which is essential for the real-world deployment for video recovery. 
DAC leverages bitstream-derived encoding priors in the embedding space as knowledge prompts, assisting effective VFM-driven video corruption detection and localization.
\item We propose a novel Corruption-aware Feature Completion module (CFC). This module effectively utilizes rich prior knowledge derived from visual foundation models, as well as self-prompting corruption embeddings, which effectively enhance the representation capacity of the completed features in existing models. The proposed DAC and CFC can be seamlessly integrated to augment current recovery methods to achieve blind recovery.
\item We conducted extensive experiments on the BSCV dataset and extended the benchmark. Experimental results show our method achieves SOTA performance in both blind and non-blind settings. Additionally, we investigated the impact and significance of this research problem on other multimedia system applications.  
\end{itemize}

\section{Related Works}
\subsection{Video Recovery}
\subsubsection{\textbf{Video Error Concealment and Inpainting}} 
Video error concealment is a standard post-processing technique used at the decoder stage to fix error regions in decoded videos~\cite{10125525}. 
Recently, deep learning methods often assume a conventional corruption pattern and employ experimental masks to mimic stripe or patch loss~\cite{sankisa2018video, xiang2019generative, 10125525, 8682097}. 
These assumptions render error concealment ineffective for handling real-world video corruption, which tends to be unpredictable and irregular.
Meanwhile, video inpainting involves generating content for unfilled video regions, utilizing masks to denote corrupted areas. 
Typical methods, such as patch-based approaches, have shown considerable success~\cite{liu2021fuseformer, zeng2020learning}. 
Currently, flow-guided generative techniques dominate video inpainting, exploiting motion to establish spatial and temporal relationships among frames~\cite{xu2019deep, kang2022error, li2022towards, zhang2024exploiting}. 
DFVI~\cite{xu2019deep} was innovative in framing video inpainting as a pixel propagation task rather than just filling RGB values. 
Li \textit{et al.}~\cite{li2022towards} improved the conventional two-stage video inpainting pipeline by optimizing it into an efficient end-to-end framework. 
ProPainter~\cite{zhou2023propainter}, a representative video inpainting method, introduces an enhanced propagation mechanism and an efficient Transformer trained in two stages, incorporating recurrent flow completion, dual-domain propagation (DDP), and a mask-guided sparse video transformer (MSVT).

For bitstream-corrupted video recovery, video inpainting is an important baseline solution because of its ability to handle arbitrary corruption. 
However, research often neglects the performance of inpainting algorithms with large and dynamic masks, which poses challenges in managing complex recovery with extensive corruption and partially preserved content from bitstream errors.

\subsubsection{\textbf{Bitstream-corrupted Video Recovery}}
Unlike the traditional corruption assumed in conventional video recovery methods, bitstream-corrupted video recovery was first proposed by Liu et al.~\cite{liu2023bitstream} in 2024.
It specifically considered communication and storage errors in real-world multimedia systems, providing a more realistic and challenging task.
The objective is to restore plausible video content in corrupted regions. 
In addition to a large-scale benchmarking dataset containing various video corruptions decoded from the corrupted video bitstream, the BSCVR was proposed to perform basic video recovery, setting a baseline.
This involves using the spatio-temporally adjacent visual information combined with any residual intact texture, color, structure, and other information in the corrupted regions. 
However, the source of masks for corruption indication is not considered, which requires manually labelled masks in practical applications. 
This is time-consuming and labor-intensive in user-involved scenarios and hinders autonomous video recovery. 
For example, real-time object detection, segmentation, stereo matching, and other applications might experience significant inference errors due to video corruption. 
Consequently, the need for manually labelled masks hampers urgent autonomous video recovery and restricts more extensive application scenarios.

\subsection{Visual Foundation Models}
Visual Foundation Models have emerged as powerful tools in computer vision, allowing a wide range of applications by learning representations that can generalize across multiple tasks. 
One notable example is the Segment Anything Model (SAM)~\cite{kirillov2023segment, ravi2024sam2}; it is designed to learn generalizable visual features and has demonstrated remarkable performance on tasks like image segmentation. 
It possesses few-shot learning capabilities, making it versatile across diverse domains.
Another example is DINO~\cite{caron2021dino, oquab2023dinov2}, which uses self-distillation to create high-quality visual representations without the need for large labeled datasets. 
DINO's ability to learn visual features through unsupervised learning has made it highly effective in various downstream tasks, including image retrieval and classification. 
Likewise, Contrastive Language-Image Pretraining (CLIP)~\cite{clip} has revolutionized multi-modal learning by jointly training image and text representations. 
Its capability to perform zero-shot classification tasks, using a large amount of unlabeled data, has established it as the cornerstone of vision-language models.
As the backbones, the Vision Transformer~\cite{dosovitskiy2020image} and its variants, such as the Data-efficient Image Transformer (DeiT)~\cite{touvron2021deit}, Hiera~\cite{ryali2023hiera}, and the Swin Transformer~\cite{liu2021swin} established the foundation of universal visual understanding. 
These models further emphasize the importance of scaling architectures and datasets to improve generalization capabilities.
In summary, the research on visual foundation models is developing in the direction of both versatility and specialization. 
By fine-tuning and adapting the foundation models, the needs of different fields and tasks can be met, promoting the widespread application of multimedia and computer vision technology.

In this paper, we propose the first blind bitstream-corrupted video recovery framework that frees the requirement of an explicit mask sequence to indicate the corrupted regions in a video.
It achieves video corruption detection and localization, which is crucial for broader application scenarios.
Based on the rich prior knowledge of the visual foundation model~\cite{clip, ravi2024sam2, caron2021dino}, we leverage bitstream-level information to prompt both effective corruption detection and enhanced content recovery.

\section{Methodology}
In this section, we present our methodology by first providing an overview of the problem setting and proposed blind bitstream-corrupted video recovery framework. 
Then, we detail the main components, highlighting their architecture and training objectives.

\subsection{Framework Overview}\label{framework_overview}
In the blind bitstream-corrupted video recovery problem, the input bitstream-corrupted frame sequence of length $L$ is denoted as $X = {\{x_i \in \mathbb{R}^{H \times W \times 3}\}}^L_{i=1}$, and its corresponding bitstream is represented by $X_b$.
The objective is to recover the frame sequence $\hat{Y} = {\{\hat{y}_i \in \mathbb{R}^{H \times W \times 3}\}}^L_{i=1}$ coherent to the corruption-free video $Y = {\{y_i \in \mathbb{R}^{H \times W \times 3}\}}^L_{i=1}$ in both the spatial and temporal domains.

As shown in Fig.~\ref{fig:teaser}, the Detect Any Corruption (DAC) module is designed to receive the input frame sequence along with its raw bitstream. 
Then it outputs 1). multi-scale visual embeddings of the video corruption extracted from the visual foundation model, and 2). localization masks to indicate the corruption regions in each frame.
Input video will be split using the estimated masks, and subsequently, be encoded.
The Corruption-aware Feature Completion (CFC) module utilizes DAC's multi-scale, cross-domain corruption embeddings and self-prompted corruption embeddings to complete and enhance the intermediate features, aiding the content recovery network in generating the recovered video

\subsection{Detect Any Corruption}
\subsubsection{\textbf{Preliminaries}}
Segment Anything Model (SAM)~\cite{kirillov2023segment} is a typical visual foundation model for prompt-based image segmentation.
Building on SAM, SAM2.1~\cite{ravi2024sam2} continues to utilize three main components: 1) a Hiera-based visual encoder pre-trained with MAE~\cite{he2022masked} as detailed in~\cite{ryali2023hiera}, 2) a prompt encoder that embeds points, bounding boxes, and masks, and 3) a mask decoder. 
It particularly expands SAM's capabilities to videos by employing a memory-based transformer for handling sequential frame processing.
Although SAM2.1 shows strong adaptability in real-world applications thanks to its pre-training on the SA-V dataset~\cite{ravi2024sam2}, it could still under-perform when dealing with video corruption that goes beyond real-world distributions.

\subsubsection{\textbf{Structure}}
\begin{figure}[t]
\centering
  \centerline{\includegraphics[width=1\linewidth]{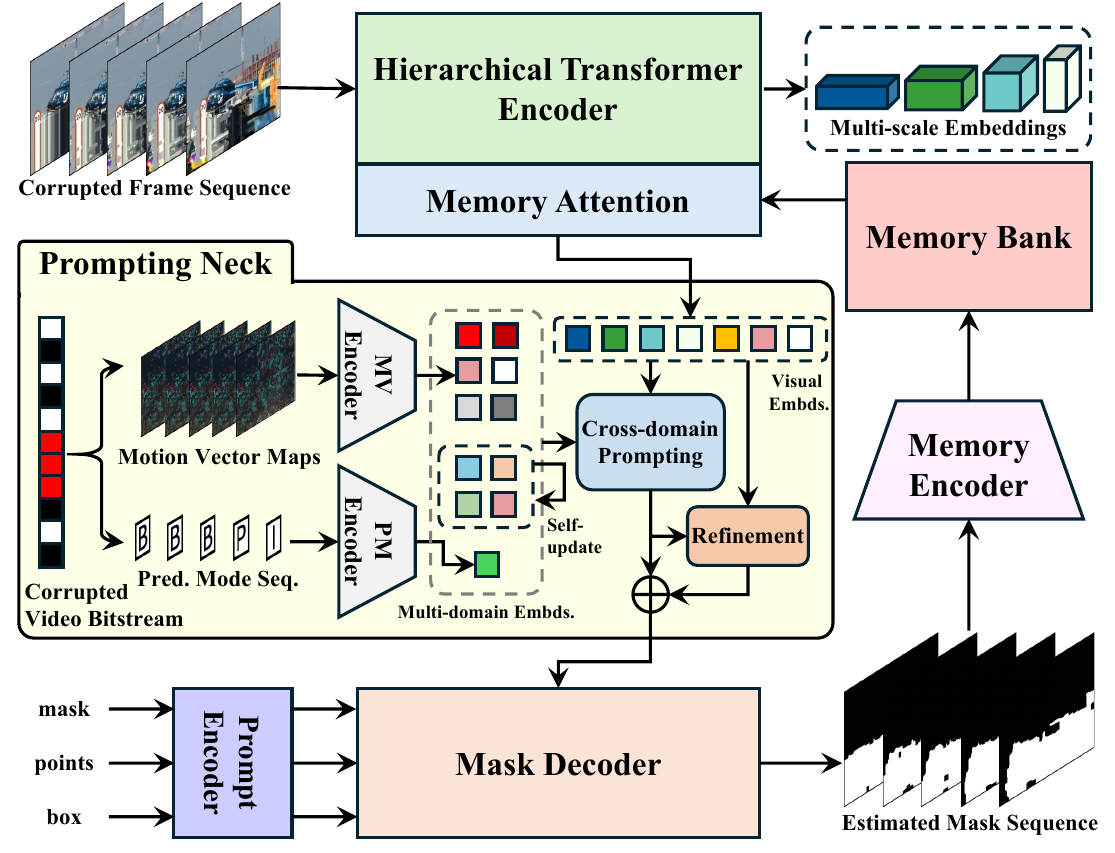}}
  \caption{Architecture of Detect Any Corruption module.}
  \vspace{-10pt}
   \label{fig:dac} 
\end{figure}

Derived from SAM's encoder-decoder framework, the Detect Any Corruption (DAC) module features an Encoder-Prompting Neck-Decoder design.
As shown in Fig.~\ref{fig:dac}, the Hiera-based~\cite{ryali2023hiera} encoder first extracts the image embeddings for each input frame.
The hierarchical structure allows the retrieval of multi-scale foundational embeddings of the input frames' content and the corruption.
Recalling memorized knowledge of previous frames stored in the memory bank, memory attention enhances the visual embedding of the current frame~\cite{ravi2024sam2}. 
Following this, the prompting neck uses the video encoding information of the current frame for further cross-domain prompting. 
As a result, the mask decoder receives the enhanced visual embeddings and different prompt embeddings encoded by the prompt encoder.
The ``two-way'' transformer-based mask decoder is applied to refine prompt and visual embeddings and generate the corresponding estimated masks, which will be further encoded into the memory bank for future enhancement in visual domain.

\subsubsection{\textbf{Cross-domain Prompting Neck}} 

Despite possessing strong open-world knowledge and zero-shot object detection capabilities, SAMs still struggle outside real-world object domains. 
Video corruption, which is unique to multimedia systems, presents challenges for SAM, causing misdetections and hindering straightforward deployment with conventional fine-tuning.
Therefore, the prompting neck is designed to address this issue and enhance SAM's ability to perceive video corruption.

\begin{figure*}[t!]
\centering
    \centerline{\includegraphics[width=1\linewidth]{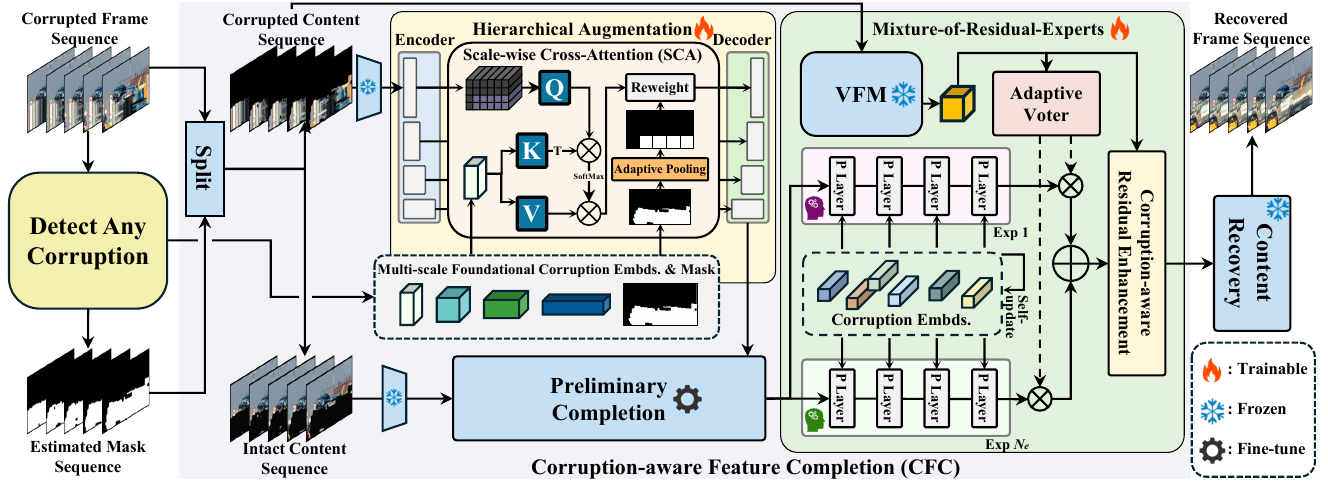}}
    \caption{Architecture of the proposed blind bitstream-corrupted video recovery framework and detailed design of the Corruption-aware Feature Completion module. During the CFC training, the trained DAC module will provide corruption indication and foundational corruption embeddings.}
    \label{fig:overview} 
    \vspace{-10pt}
\end{figure*}

For each local input image denoted as $x_m$, to identify the correlation of each frame, we derive the corresponding prediction mode from the standard ffmpeg video codec and transform it into a one-hot vector $p_m$.
Additionally, the motion vectors extracted for each eligible frame from the video bitstreams will be denoted $v_m$.
Then we map the motion vectors' direction to the HSV color space and fill the corresponding macroblocks to obtain the motion vector map. 
Blending it with the corresponding frame in a certain ratio $\eta$, the motion vector map can highlight areas related to inter-frame dependencies and compression encoding, which can be applied to guide successive detection and localization.
As shown in Fig.~\ref{fig:overview}, the input includes 1). $X_{in} = {\{x_m \in \mathbb{R}^{H \times W \times 3}\}}^{N}_{m=1}$ which is a frame sequence with a length of $N=N_l + N_{nl}$, consisting of $N_l$ local frames to be restored and $N_{nl}$ non-local reference frames sampled from the rest of the video, 2). the corresponding motion vector map sequence $V_{in} = {\{v_m \in \mathbb{R}^{H \times W \times 3}\}}^{N_l}_{m=1}$, and 3). the prediction mode sequence $P_{in} = {\{p_m \in \mathbb{R}^{1 \times 3}\}}^{N_l}_{m=1}$.
We first tokenize the input $V_{in}$ and $P_{in}$ via two encoders, embedding the video coding information into prompt tokens.
Since motion vector maps have not been considered in prior training of visual foundation models, we employ the DINOv2 encoder, pre-trained through self-supervised learning, as the motion vector encoder. 
DINOv2 excels at extracting robust visual features and capturing local and global relationships, thereby facilitating effective prompt embeddings~\cite{oquab2023dinov2} when incorporated with an MLP-based adaptation block.
Additionally, a prediction mode tokenizer is also introduced to encode the one-hot vector input into a prediction mode prompt token. 
Finally, by combining $N_p$ self-updated corruption prompts, we can construct a multi-domain prompt pool.

To achieve effective cross-domain prompting, we first project the foundational visual embeddings $\textbf{F}^{\text{f}}$ into query tokens and generate key and value tokens using the multi-domain prompts. 
Then we apply the Token-Dictionary Cross Attention mechanism~\cite{zhang2024atd}, expressed as $\textbf{F}^{\text{f}'} = \text{SoftMax} \left(\frac{\text{Sim}_{\cos} (\mathbf{Q}, \mathbf{K})}{\tau} \right)\mathbf{V}$, which computes a cosine similarity map between normalized Query and Key, using a learnable scaling parameter $\tau$ and SoftMax function; the attention map can be obtained and produce attention-weighted Value. 
We further employ self-attention to the foundational embeddings $\textbf{F}^{\text{f}}$ and enhanced embeddings classified by different prompt attention maps. 
The combination of refined embeddings are denoted as $\hat{\textbf{F}}^{\text{f}}$, from which a corruption-indicating mask sequence can be decoded.

\subsection{Corruption-aware Feature Completion}
\subsubsection{\textbf{Observation}}
As illustrated in Fig.~\ref{fig:teaser}, the volume of useful information directly impacts the feature completion quality. 
However, the existing feature completion module exhibits blind inclusion of interfering artifacts, such as misaligned structure and distorted color, which do not always positively contribute to the recovery process. 
The interfering corruption feature will be embedded in the intermediate feature used for content recovery, which will ultimately limit the content recovery performance.

\subsubsection{\textbf{Structure}}
To achieve improved residual mining and balancing, we propose the Corruption-aware Feature Completion (CFC) module. 
It aims to augment the residual embeddings within the corruption features $\textbf{F}^{\text{c}}$ by the multi-scale foundational corruption embeddings provided by DAC.
As shown in Fig.~\ref{fig:overview}, $N_e$ residual experts $\text{Exp}_i(\cdot)$ are applied to collaborate in a mixture-of-experts (MoE) architecture to perform fine-grained feature recovery. 
For each expert, recovery is directed by shared corruption tokens that are learned through the network's self-prompting.
Then, a VFM is deployed to perform high-level corruption understanding, which is used in 1). vote for the contribution of each expert and 2). enhance the informative residual information in a channel-wise manner. 
This design enables noticeable feature quality improvement, resulting in more faithful recovery performance when integrating with the pre-trained content recovery network.  

\subsubsection{\textbf{Hierarchical Corruption Feature Augmentation}}
Specifically, the DAC-estimated mask sequence first split the input video into the intact content sequence and the corrupted content, denoted as $X^{\text{i}}_{in} = {\{x_m \odot (1-m_m)\}}^{N}_{m=1}$ and $X^{\text{c}}_{in} = {\{x_m \odot m_m\}}^{N}_{m=1}$.
In addition, the mentioned multi-scale foundational corruption embedding $\hat{\textbf{F}}^{\text{f}}=\{\hat{\textbf{F}}^f_j \in \mathbb{R}^{H_j \times W_j \times C_j}\}^{S}_{j=1}$ is applied to perform hierarchical augmentation to the basic corruption features $\textbf{F}^{c}$. 
The block adopts a convolutional U-shaped~\cite{ronneberger2015u} encoder-decoder structure and leverages a scale-wise cross-attention (SCA) mechanism to facilitate effective feature augmentation at different scales.
In detail, given the corruption feature on scale $j$, denoted as $\textbf{F}^c_j$, the foundational embedding $\hat{\textbf{F}}^{\text{f}}_j$, and the corruption mask $\textbf{M}$, SCA first applies reduced $Q,K,V$ projection and the attention scores are computed, then the attended feature representation is re-weighted by the interpolated mask $\textbf{M}'$. This process can be represented as:
\begin{equation}
    \textbf{F}'^c_j = \text{SoftMax} \left( QK^\top/\sqrt{d} \right)V \odot \delta\textbf{M}'
\end{equation}
where $\odot$ denotes element-wise multiplication, $\delta$ represents the reweighting intensity.  
The augmented features will be reconstructed by the decoder with skip connections and a learnable residual weighting mechanism by $\hat{\textbf{F}^c_j} = \lambda \cdot \textbf{F}'^c_j + (1 - \lambda) \cdot \textbf{F}^c_j$, where $\lambda$ is a learnable parameter that adaptively balances the augmentation. 
This design enables hierarchical augmentation of the corruption features while maintaining spatial coherence, making it more reliable for subsequent feature completion.  

\definecolor{lightgray}{gray}{0.95}
\renewcommand{\arraystretch}{1.1}
\begin{table*}
\setlength{\tabcolsep}{10pt}
\centering
\caption{Comparison of different methods on BSCV~\cite{liu2023bitstream} subsets. * denotes BSCV-retrained baseline. A BSCV-finetuned SAM2.1 is used to extend non-blind baselines. Non-blind evaluation is conducted using BSCV oracle masks without applying detectors.}
\label{tab:blind}
\begin{tabular}{lccccccccc} 
\toprule
\multirow{2}{*}{Methods} & \multirow{2}{*}{Detector} & \multicolumn{4}{c}{YouTube-VOS~\cite{vos2019} Subset}                               & \multicolumn{4}{c}{DAVIS~\cite{Pont-Tuset_arXiv_2017} Subset}                                      \\ 
\cline{3-10}
                                                    &                           & PSNR           & SSIM            & LPIPS           & VFID            & PSNR           & SSIM            & LPIPS           & VFID             \\ 
\hline  
STTN~\cite{zeng2020learning}                        & \multirow{8}{*}{SAM2.1}      & 28.50          & 0.9019          & 0.0504          & 0.0386          & 22.96          & 0.7509          & 0.1004          & 0.2858           \\
STTN*~\cite{zeng2020learning}                       &                           & 29.13          & 0.9152          & 0.0452          & 0.0368          & 23.34          & 0.7669          & 0.0951          & 0.2815           \\
FuseFormer~\cite{liu2021fuseformer}                 &                           & 27.75          & 0.8850          & 0.0608          & 0.0458          & 22.85          & 0.7521          & 0.1022          & 0.2949           \\
FuseFormer*~\cite{liu2021fuseformer}                &                           & 28.87          & 0.9150          & 0.0456          & 0.0373          & 23.24          & 0.7726          & 0.0972          & 0.2783           \\
E2FGVI-HQ~\cite{li2022towards}                      &                           & 28.81          & 0.9079          & 0.0509          & 0.0390          & 23.59          & 0.7809          & 0.0916          & 0.2774           \\
E2FGVI-HQ*~\cite{li2022towards}                     &                           & 30.12          & 0.9335          & 0.0377          & 0.0333          & 24.43          & 0.8125          & 0.0812          & 0.2632           \\
ProPainter~\cite{zhou2023propainter}                &                           & 29.81          & 0.9302          & 0.0390          & 0.0521          & 24.26          & 0.8103          & 0.0793          & 0.2113           \\
BSCVR-P*~\cite{liu2023bitstream}                    &                           & \uline{30.71}  & \uline{0.9432}  & \uline{0.0309}   & \uline{0.0306}   & \uline{26.41}   & \uline{0.8896}   & \uline{0.0499}          & \uline{0.1998}           \\ 
Ours*                                               & DAC                       & \textbf{31.60} & \textbf{0.9535} & \textbf{0.0297} & \textbf{0.0292} & \textbf{28.14} & \textbf{0.9086} & \textbf{0.0434} & \textbf{0.1520}  \\
\hline
STTN~\cite{zeng2020learning}                        & \multirow{9}{*}{Oracle}     & 29.30          & 0.9150          & 0.0472          & 0.0370          & 26.21          & 0.8603          & 0.0634          & 0.1868           \\
STTN*~\cite{zeng2020learning}                       &                                & 29.94          & 0.9283          & 0.0420          & 0.0350          & 26.61          & 0.8731          & 0.0584          & 0.1808           \\
FuseFormer~\cite{liu2021fuseformer}                 &                                & 28.74          & 0.9010          & 0.0556          & 0.0411          & 26.19          & 0.8592          & 0.0662          & 0.1933           \\
FuseFormer*~\cite{liu2021fuseformer}                &                                & 29.77          & 0.9284          & 0.0421          & 0.0351          & 26.71          & 0.8801          & 0.0557          & 0.1761           \\
E2FGVI-HQ~\cite{li2022towards}                      &                                & 29.69          & 0.9228          & 0.0469          & 0.0363          & 26.79          & 0.8765          & 0.0600          & 0.1806           \\
E2FGVI-HQ*~\cite{li2022towards}                     &                                & 31.00          & 0.9473          & 0.0342          & 0.0315          & 27.66          & 0.9018          & 0.0491          & 0.1646           \\
ProPainter~\cite{zhou2023propainter}                &                                & 30.40          & 0.9389          & 0.0369          & 0.0497          & 27.24          & 0.8920          & 0.0513          & 0.1389           \\
BSCVR-P*~\cite{liu2023bitstream}                    &                                & \uline{31.56}  & \uline{0.9536}  & \uline{0.0288}  & \uline{0.0296}  & \uline{28.29}  & \uline{0.9147}  & \textbf{0.0395} & \uline{0.1484}   \\
Ours*                                               &                                & \textbf{32.03} & \textbf{0.9605} & \textbf{0.0279} & \textbf{0.0286} & \textbf{28.60} & \textbf{0.9207} & \uline{0.0403}  & \textbf{0.1461}  \\
\bottomrule
\end{tabular}
\vspace{-1em}
\end{table*}

\subsubsection{\textbf{Mixture-of-Residual-Experts}}
In the feature completion process, we first apply the feature completion module of BSCVR to obtain a preliminary feature $\textbf{F}_{\text{b}}$, which is input to the Mixture-of-Residual-Experts (MoRE) block.
This mixture-of-experts (MoE) structure is coordinated by a high-level understanding of corruption patterns without explicit supervision. 
To achieve this, we propose to take advantage of open-world knowledge from the CLIP encoder~\cite{clip}, from which an implicit representation of the video corruption pattern can be obtained as a token denoted as $\textbf{F}^c$.
In the adaptive voter, an MLP-based adaptation is applied to transform the extracted token $\textbf{F}^c$ into adapted corruption embeddings by $\textbf{F}^c_{\text{CLIP}} = \text{MLP}(\text{CLIP}_{\text{enc}}(\textbf{F}^c))$.
Therefore, the adapted high-level corruption embeddings can serve as a reference point for the soft voting gate to generate a weighting vector $\textbf{w}$ to coordinate the residual experts:
\begin{gather}
        \textbf{w} = \text{SoftMax}(\text{Voter}(\textbf{F}^c_{\text{CLIP}})) \in \mathbb{R}^{1 \times N_e}\\
        \textbf{F}'^c_{\text{refined}} = \sum_i w_i \cdot \text{Exp}_i(\textbf{F}_{\text{b}}, \textbf{P}), \quad \text{Exp}_i \in \{\text{Exp}_1, \ldots, \text{Exp}_{N_e}\}.
\end{gather}
$\textbf{w}$ is reshaped to match the spatial dimensions of the feature maps and used to dynamically combine the output of $N_e$ experts. 
This process is realized by each expert focusing on different residual mining; each cross-attention-based P-layer can adaptively leverage the self-update corruption embeddings $\textbf{P}$ to prompt the corresponding feature recovery.
This Mixture-of-Residual-Experts (MoRE) structure enables dynamical coordination to combine the experts' outputs based on high-level corruption understanding, ensuring robustness against various types of corruption with different degrees.  

To further emphasize the interaction of the features across different levels, we employ a corruption-aware residual enhancement guided by a high-level understanding of corruption. 
Given two input embeddings, $\textbf{F}^c_{\text{refined}}$ and $\textbf{F}^c_{\text{CLIP}}$, the query features $Q_\text{C}$ are derived from $\textbf{F}'^c_{\text{CLIP}}$ using a channel pooling operation, while $K_\text{R}$ and $V_\text{R}$ are extracted from $\textbf{F}^c_{\text{refined}}$ using global average pooling (GAP).  
Subsequently, the resulting cross-attention scores can modulate the importance of $\textbf{F}^c_{\text{refined}}$ channels through a learnable MLP and activation of the sigmoid, generating adaptive channel weights. 
By channel-wise multiplication, dynamic re-weighting of channels based on cross-input dependencies can be achieved, the simplified representation is shown below:
\begin{equation}
    \hat{\textbf{F}}^c_{\text{refined}} = \textbf{F}'^c_{\text{refined}} \cdot \text{Sigmoid}(\text{MLP}(\text{CA}(Q_\text{C}, K_\text{R}, V_\text{R}))),
\end{equation}
where $\text{CA}(\cdot)$ denotes the cross-attention calculation.

As a result, our CFC module can achieve 1). effective augmentation to the corruption feature, 2). dynamical and corruption-aware completion to generate an enhanced intermediate feature that can collaborate better with the pre-trained content recovery network, leading to more credible outcomes in recovery.

\section{Experiments}

\subsection{Experimental Settings} \label{setting}

\begin{figure*}[t]
  \includegraphics[width=1\linewidth, height=10cm]{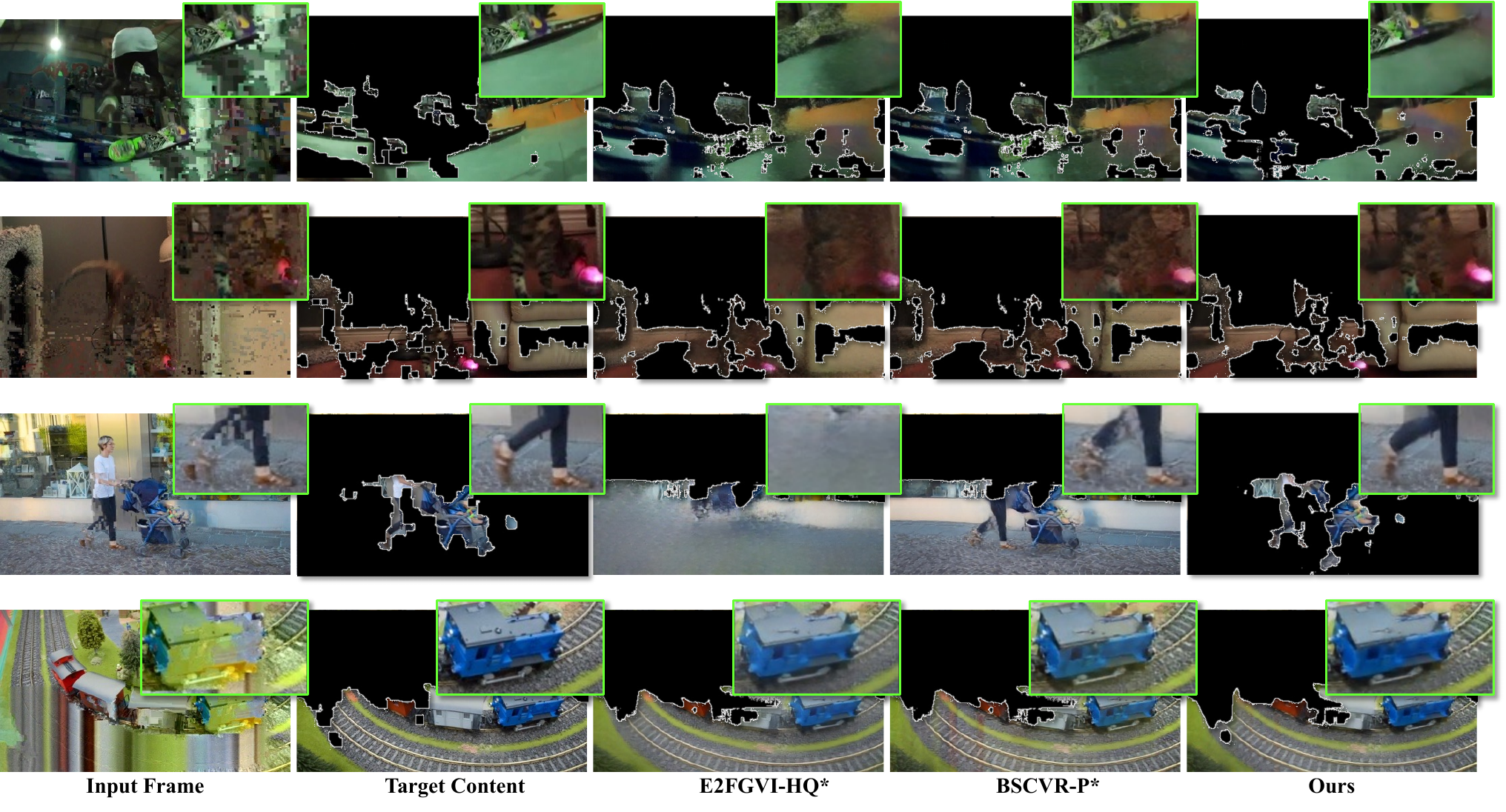}
  \vspace{-20pt}
  \caption{Visual comparison results under blind recovery setting. We visualize the masked regions indicated by the SAM2.1 and DAC to simultaneously demonstrate corruption detection and video corruption recovery performance within the target areas.}
  \label{fig:qual-eval}
\end{figure*}

In the experiment, we first train our DAC model using the main branch of the BSCV training set~\cite{liu2023bitstream} for 50 epochs, using the AdamW optimizer, with a batch size of 2, a frame number of 5, and a basic learning rate set to 5e-5. 
The Hiera encoder, the prompt encoder, and the mask decoder of the DAC were initialized using the publicly available SAM2.1-tiny checkpoint pre-trained on the SA-V dataset~\cite{ravi2024sam2}.
The CFC training runs for 100K iterations using the Adam optimizer with a batch size of 4, with a local frame number $N_l$ set to 5 and a non-local frame number $N_{nl}$ set to 3.
The resolution is set to $432\times240$ to reduce the training cost and the overall learning rate is set to 1e-4, while the BSCVR feature completion module is fine-tuned with a learning rate of 1e-5 and the content recovery module is frozen. 
For baselines, we used the fine-tuned SAM2.1-tiny model trained with the officially recommended configuration.
The predicted masks will replace the BSCV-provided oracle masks and be sent into existing recovery models to perform basic blind recovery.
The whole framework is trained in a two-stage manner; the DAC training is supervised using the SAM2.1 official combination of focal loss, dice loss, L1 loss, and cross-entropy loss~\cite{ravi2024sam2}; the CFC module is trained according to the BSCVR setting~\cite{liu2023bitstream}.
Training and evaluation are conducted with a 4 NVIDIA RTX 3090 GPUs workstation. Codes will be available at this \href{https://github.com/LIUTIGHE/B2SCVR}{\textcolor{magenta}{link}}.

\subsection{Quantitative Evaluation}
We utilize several metrics to assess the quality of the spatial frame and the temporal consistency of the recovery output. 
PSNR and SSIM~\cite{wang2004image} are used to assess the similarity between frames and actual data, while perceptual similarity, temporal continuity, and smoothness are evaluated with VFID~\cite{kim2019deep} and LPIPS~\cite{zhang2018unreasonable}.
In the blind recovery scenario, we comprehensively compared the performance of our proposed framework with other existing methods. 
Among them, the existing non-blind recovery method, as mentioned earlier, uses the fine-tuned SAM2.1 as the corruption detector.
Unlike the original benchmark in BSCV~\cite{liu2023bitstream}, which is evaluated on a sampled test set with a sampling rate $1/5$, our setting avoids the potential negative impact on baseline methods caused by reduced referable spatio-temporal information.
In our extended benchmark, we performed the comparison based on the whole test set.
The experimental results show that our method consistently outperforms existing methods in all metrics and on two subsets of BSCV, achieving new state-of-the-art results. 
Compared to the best combination, BSCVR-P and SAM2.1, our method has a significant improvement in reconstruction quality regarding PSNR and SSIM. 
Additionally, realism is also markedly improved; on the YouTube-VOS subset, our method increases PSNR by 0.89dB, and on the DAVIS subset, PSNR is improved by 1.73dB, with VFID reduced by 0.0478.
It is worth noting that although ProPainter achieves the best performance among the pre-trained models (i.e., those without * marks), its two-stage design and 40\% longer training schedule make both training and deployment considerably more challenging. Given the lack of significant performance advantage, we did not re-train it on BSCV. Based on the observed trend, it is unlikely to outperform the other baselines under the same training settings.

\subsection{Qualitative Evaluation}
We also performed a qualitative evaluation and, as shown in Fig.~\ref{fig:qual-eval}.
We present the corrupted input video frames alongside the designated target contents marked by the oracle mask provided by BSCV in columns 1 and 2. 
The predicted region requiring recovery given by different methods and the corresponding restoration results are also demonstrated.
From our observation, our method has a stronger capability to locate the corruption, avoiding redundant corruption indication, thus reducing the area and difficulty of recovery. 
By increasing the available neighboring information, our method, as seen in the first row, avoids faulty recovery for the skateboard pattern but focuses on the background environment repair, thereby reducing block artifacts.
In the fourth line, it can be observed that even though the target region is similar, our method can more accurately recover the color of the toy locomotive and reduce interfering color stripes, demonstrating superior performance.

\subsection{Ablation Studies}
We conducted extensive ablation experiments to demonstrate the contributions of the DAC and CFC modules and their components.

\begin{table}[hbt!]
\centering
\caption{Evaluation and ablation study results of DAC module on BSCV dataset, * denotes fine-tuning following SAM2.1~\cite{ravi2024sam2} officially recommended configuration.}
\label{tab:dac-ab}
\begin{tabular}{lcccccc} 
\toprule
\multirow{2}{*}{Methods} & \multicolumn{2}{c}{Prompts}                             & \multirow{2}{*}{\begin{tabular}[c]{@{}c@{}}Mean \\IoU\end{tabular}} & \multirow{2}{*}{\begin{tabular}[c]{@{}c@{}}Mean \\Dice\end{tabular}} & \multirow{2}{*}{\begin{tabular}[c]{@{}c@{}}Mean \\Acc\end{tabular}} & \multirow{2}{*}{\begin{tabular}[c]{@{}c@{}}Mean \\Recall\end{tabular}}  \\ 
\cline{2-3}
                         & Deg.                     & Bits.         &                       &                     &                 &                   \\ 
\hline
SAM2.1*                  & \ding{56}                & \ding{56}     & 0.55                  & 0.66                & 0.93            & 0.75              \\
SAM2.1                   & \ding{56}                & \ding{56}     & 0.62                  & 0.73                & 0.96            & 0.65              \\ 
\hline
DAC- -                   & \ding{52}                & \ding{56}     & 0.63                  & 0.74                & 0.96            & 0.64              \\
DAC-                     & \ding{52}                & MLP           & 0.63                  & 0.73                & 0.96            & 0.73              \\
DAC                      & \ding{52}                & DINOv2        & \textbf{ 0.63}        & \textbf{0.74}       & \textbf{0.96}   & \textbf{0.78}     \\
\bottomrule
\end{tabular}
\vspace{-10pt}
\end{table}

\subsubsection{\textbf{DAC Module}}

We first evaluated the fine-tuned SAM2.1 and conducted a series of enhancement experiments based on it. The fine-tuned model (SAM2.1*) performed poorly in terms of Mean IoU (0.55) and mean dice (0.66), indicating a limited generalizability for this task. In contrast, the re-trained SAM2.1 achieved higher IoU (+0.07) and Dice (+0.07), but exhibited a notable drop in Mean Recall (-0.10), which is critical since corruption detection should be sensitive to false negatives.
We also observed that using corruption prompts alone (DAC--) failed to effectively mitigate the low recall issue. 
Then we introduced the bitsrteam prompts, and it can be observed that the performance across metrics is more balanced.
Furthermore, we explored the effectiveness using DINOv2, which shows that pre-trained VFM can consistently outperform MLP, achieving the best results across all metrics. 
These results demonstrate that the use of VFM's open-world knowledge effectively improves the detection and localization performances.

\begin{table}[hbt!]
\centering
\caption{Ablation study results of the two main components, MoRE and HA, in the CFC module.}
\label{tab:dfc-ab1}
\begin{tabular}{lccccccc} 
\toprule
\multirow{2}{*}{Methods} & \multicolumn{2}{c}{Components}    & \multirow{2}{*}{PSNR} & \multirow{2}{*}{SSIM} & \multirow{2}{*}{LPIPS} & \multirow{2}{*}{VFID}   \\ 
\cline{2-3}
                & MoRE              & HA                     &                       &                       &                        &                         \\ 
\hline
Baseline        & \ding{56}         & \ding{56}              & 28.29                 & 0.9147                & 0.0395                 & 0.1482                  \\ 
CFC-            & \ding{52}         & \ding{56}              & 28.57                 & 0.9202                & 0.0397                 & \textbf{0.1460}         \\
CFC             & \ding{52}         & \ding{52}              & \textbf{28.60}       & \textbf{0.9208}        & \textbf{0.0414}        & 0.1476                  \\
\bottomrule
\end{tabular}
\end{table}

\subsubsection{\textbf{CFC Module}}
To demonstrate CFC’s effectiveness, in Tab.~\ref{tab:blind}, we conducted detailed BSCV non-blind recovery benchmarking on the full test set using pre-defined oracle masks from BSCV, avoiding unfair comparisons caused by DAC’s precise indications. Our method consistently outperforms others. Compared to BSCVR-P, it improves reconstruction quality and perceptual realism, with PSNR gains of 0.4\textasciitilde0.5dB and SSIM improvements of 0.06\textasciitilde0.07, indicating better feature completion using external knowledge.

We also performed a detailed ablation study for the CFC to justify its design. 
As shown in Table~\ref{tab:dfc-ab1}, it can be seen that the proposed mixture-of-residual-experts (MoRE) can collaborate effectively and enhance recovery performance.
Also, the hierarchical feature augmentation (HA) using DAC-provided multi-scale embeddings also benefits the recovery performance. 

\begin{table}[hbt!]
\centering
\caption{Ablation study results for MoRE in CFC module}
\label{tab:dfc-ab2}
\begin{tabular}{lccccccc} 
\toprule
\multirow{2}{*}{Methods} & \multicolumn{2}{c}{Components}                                       & \multirow{2}{*}{PSNR} & \multirow{2}{*}{SSIM}   & \multirow{2}{*}{LPIPS} & \multirow{2}{*}{VFID}\\ 
\cline{2-3}
                         & MoRE                                     & HA                       &                       &                         &                         &                    \\ 
\hline
Baseline                 & \ding{56}                                & \ding{56}                & 28.29                 & 0.915                  & 0.040                 & 0.148                \\ 
CFC- -                   & Linear/2E                                & \ding{56}                & 28.46                 & 0.919                  & 0.040                 & 0.147                \\
CFC-                     & VFM/2E                                   & \ding{56}                & 28.57                 & 0.920                  & 0.040                 & \textbf{0.146}       \\
\hline
CFC\_e1                  & -/1E                                     & \ding{52}                & 28.39                 & 0.917                  & 0.041                 & 0.149                \\
CFC                      & VFM/2E                & \ding{52}        & \textbf{ 28.60}          & \textbf{0.921}        & \textbf{0.041}         & 0.148                                        \\
CFC\_e3                  & VFM/3E                                   & \ding{52}                & 28.54                 & 0.919                  & 0.041                 & 0.147                \\
\bottomrule
\end{tabular}
\end{table}

Based on the results of the additional ablation study in Fig.~\ref{tab:dfc-ab2} for the MoE structure design analysis, we demonstrate the superiority of VFM-based coordination, compared to the vanilla linear gating mechanism~\cite{jacobs1991adaptive}.
It achieves improved recovery performance and coordination among residual experts, and it can be found that adding more residual experts has little effect; to some extent, it aligns with our observation that structure and color are the two main crucial residuals for video feature completion.

\section{Significance}
Additionally, we demonstrate the importance of blind bitstream-corrupted video recovery for different multimedia systems. 
Video corruption not only degrades visual quality and user experience but also introduces functional risks to downstream tasks. 
For example, object detection and multi-modal understanding are fundamental to intelligent video analysis: the former identifies and localizes visual entities, while the latter (e.g., captioning) generates semantic interpretations. 
Both rely on high-quality visual input for reliable performance.
In the qualitative evaluation, it can be observed that video corruption may disturb the performance of object detection using YOLOv11~\cite{khanam2024yolov11} and GPT-4o~\cite{hurst2024gpt}-based visual captioning.
It can be observed that the detection of the paraglider and motorcycles is failed and the GPT model cannot accurately describe the contents.
In contrast, our recovered results can help improve the reliability, enabling consistent multi-modal analysis. 
This demonstrates the practical necessity of blind bitstream-corrupted video recovery to maintain the integrity of real-world multimedia systems.

\begin{figure}[hbt!]
    \centering
    \includegraphics[width=1\linewidth]{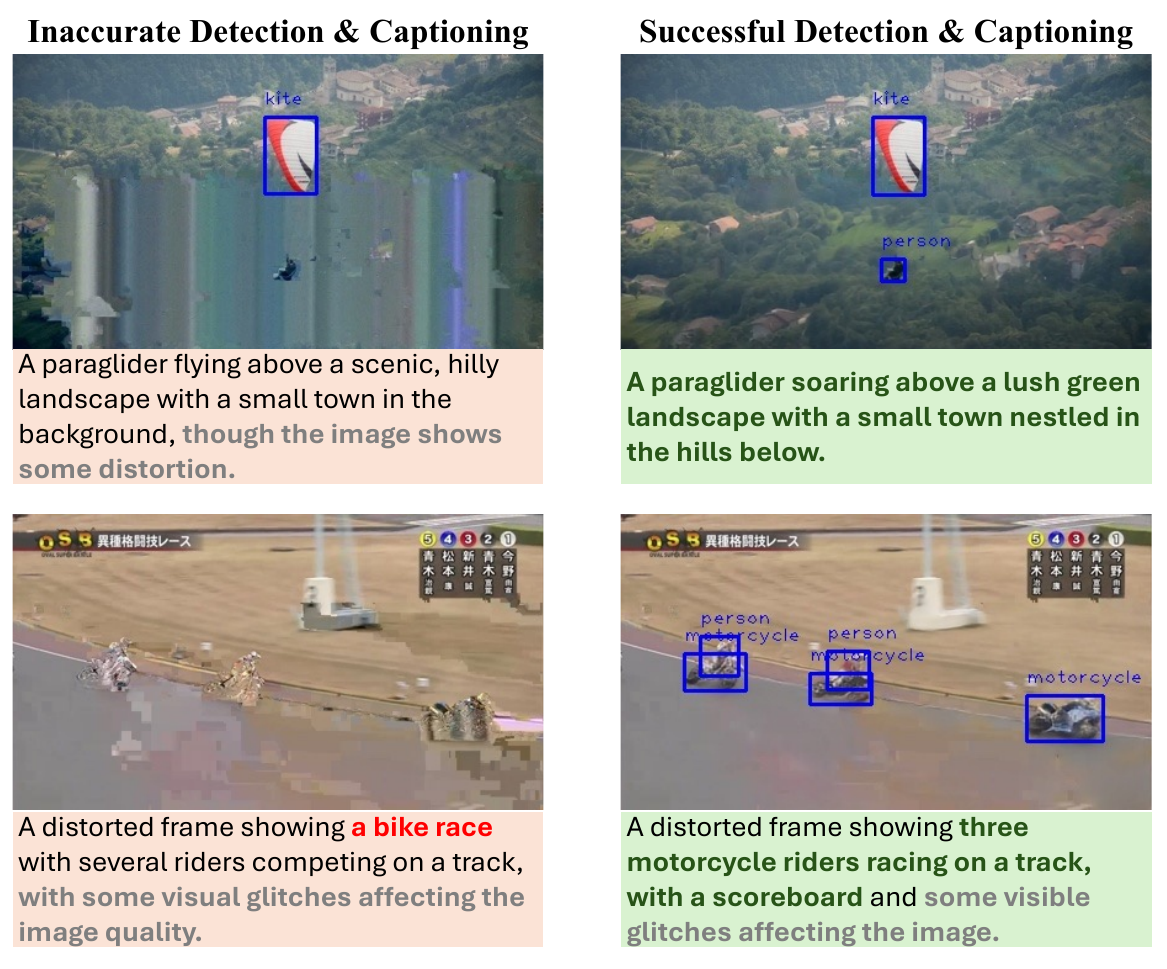}
    \caption{Recovering bitstream-corrupted video improves the resilience of multimedia systems in noisy environments, mitigating the risk of perceptual errors.}
    \vspace{-10pt}
    \label{fig:appl}
\end{figure}

\section{Conclusion}
This paper presents a novel blind bitstream-corrupted video recovery framework that effectively addresses the challenges associated with video corruption detection and disruptive residual processing. 
By integrating visual foundation models with a specialized recovery model, the proposed approach successfully adapts to various corruption and video coding information. 
The DAC model leverages rich visual priors and bitstream knowledge of video coding to enhance corruption localization and recovery, while the CFC module intelligently processes residual contributions to suppress artifacts and highlight informative details. 
Experimental results demonstrate the superior performance of our method in recovering bitstream-corrupted videos, offering a state-of-the-art solution that eliminates the need for labor-intensive annotations. 
The proposed framework paves the way for improved multimedia communication and storage systems, ensuring enhanced user experience, integrity of multi-modal perception, and broader applicability.

\bibliographystyle{ACM-Reference-Format}
\balance
\bibliography{B2SCVR_ref}

\end{document}